\documentclass[twocolumn,showpacs,preprintnumbers,amsmath,amssymb,aps,pra,superscriptaddress]{revtex4}

\usepackage{graphicx}% Include figure files
\usepackage[tight]{units} %helps writing units correctly
\usepackage{textcomp} %units and stuff
\usepackage{gensymb} %units and stuff

\begin{document}

\title{EIT ground-state cooling of long ion strings}
%\title{Electromagnetically-induced-transparency cooling of long ion strings}% Force line breaks with \\

\author{Regina Lechner}
\affiliation{Institut f\"ur Quantenoptik und Quanteninformation, \"Osterreichische Akademie der Wissenschaften,
Technikerstra\ss{}e 21a, 6020 Innsbruck, Austria}
\affiliation{Institut f\"ur Experimentalphysik, Universit\"at Innsbruck, Technikerstra\ss{}e 25, 6020 Innsbruck, Austria}

\author{Christine Maier}
 \affiliation{Institut f\"ur Quantenoptik und Quanteninformation, \"Osterreichische Akademie der Wissenschaften,
Technikerstra\ss{}e 21a, 6020 Innsbruck, Austria}
\affiliation{Institut f\"ur Experimentalphysik, Universit\"at Innsbruck, Technikerstra\ss{}e 25, 6020 Innsbruck, Austria}

\author{Cornelius Hempel\footnote{Current address: ARC Centre for Engineered Quantum Systems, School of Physics, The University of Sydney, NSW 2006 Australia}}
 \affiliation{Institut f\"ur Quantenoptik und Quanteninformation, \"Osterreichische Akademie der Wissenschaften,
Technikerstra\ss{}e 21a, 6020 Innsbruck, Austria}
\affiliation{Institut f\"ur Experimentalphysik, Universit\"at Innsbruck, Technikerstra\ss{}e 25, 6020 Innsbruck, Austria}

\author{Petar Jurcevic}
 \affiliation{Institut f\"ur Quantenoptik und Quanteninformation, \"Osterreichische Akademie der Wissenschaften,
Technikerstra\ss{}e 21a, 6020 Innsbruck, Austria}
\affiliation{Institut f\"ur Experimentalphysik, Universit\"at Innsbruck, Technikerstra\ss{}e 25, 6020 Innsbruck, Austria}

\author{Ben P. Lanyon}
\affiliation{Institut f\"ur Quantenoptik und Quanteninformation, \"Osterreichische Akademie der Wissenschaften,
Technikerstra\ss{}e 21a, 6020 Innsbruck, Austria}
\affiliation{Institut f\"ur Experimentalphysik, Universit\"at Innsbruck, Technikerstra\ss{}e 25, 6020 Innsbruck, Austria}

\author{Thomas Monz}
\affiliation{Institut f\"ur Experimentalphysik, Universit\"at Innsbruck, Technikerstra\ss{}e 25, 6020 Innsbruck, Austria}

\author{Michael Brownnutt\footnote{Current address: The University of Hong Kong, Pok Fu Lam, Hong Kong}}
\affiliation{Institut f\"ur Experimentalphysik, Universit\"at Innsbruck, Technikerstra\ss{}e 25, 6020 Innsbruck, Austria}

\author{Rainer Blatt}
\affiliation{Institut f\"ur Quantenoptik und Quanteninformation, \"Osterreichische Akademie der Wissenschaften,
Technikerstra\ss{}e 21a, 6020 Innsbruck, Austria}
\affiliation{Institut f\"ur Experimentalphysik, Universit\"at Innsbruck, Technikerstra\ss{}e 25, 6020 Innsbruck, Austria}

\author{Christian F. Roos\footnote{christian.roos@uibk.ac.at}}
\affiliation{Institut f\"ur Quantenoptik und Quanteninformation, \"Osterreichische Akademie der Wissenschaften,
Technikerstra\ss{}e 21a, 6020 Innsbruck, Austria}
\affiliation{Institut f\"ur Experimentalphysik, Universit\"at Innsbruck, Technikerstra\ss{}e 25, 6020 Innsbruck, Austria}

%\email{christian.roos@uibk.ac.at}

\date{\today}

\begin{abstract}
Electromagnetically-induced-transparency (EIT) cooling is a ground-state cooling technique for trapped particles. EIT offers a broader cooling range in frequency space compared to more established methods. In this work, we experimentally investigate EIT cooling in strings of trapped atomic ions. In strings of up to 18 ions, we demonstrate simultaneous ground state cooling of all radial modes in under 1 ms. This is a particularly important capability in view of emerging quantum simulation experiments with large numbers of trapped ions. Our analysis of the EIT cooling dynamics is based on a novel technique enabling single-shot measurements of phonon numbers, by rapid adiabatic passage on a vibrational sideband of a narrow transition.
\end{abstract}

%\pacs{37.10.Jk, 42.50.Gy, 03.67.Ly}% PACS, the Physics and Astronomy Classification Scheme.

\maketitle

\section{\label{sec:intro}Introduction}
The development of laser cooling \cite{Haensch1975,Wineland1975} has resulted in the preparation of neutral atoms \cite{Phillips1998,Cohen-Tannoudji1998} and ions \cite{Wineland1978,Neuhauser1978} at temperatures ranging from milli- to micro-kelvin or even below. Cooling trapped particles to low temperatures enables the spatial localization of single particles at length scales which can be much smaller than optical wavelengths, as well as the preparation of dense cold samples of atoms. This excellent quantum control over the motional degrees of freedom of atoms and ions has led to the wide-spread use of laser cooling techniques in various research fields, such as atomic clocks \cite{Gibble1993,Ludlow2015}, atom interferometry \cite{Cronin2009}, ultra-cold quantum gases \cite{Bloch2008}, and quantum information processing \cite{Haeffner2008,Saffman2010} with atoms and ions. 

While the minimum temperature achievable by a laser cooling technique might be considered to be the ultimate figure of merit, it is by no means the only one. Other characteristics of a laser cooling scheme are the range of initial temperatures over which it is applicable, the achievable cooling rate and, in the case of trapped particles, the range of trap oscillation frequencies over which the cooling scheme is effective. 
%For this reason, Doppler cooling \cite{Haensch1975,Wineland1978,Neuhauser1978} has not remained the only scheme. More 
For this reason, sophisticated schemes using interference of several laser beams and/or multiple internal electronic states \cite{Diedrich1989,Wineland1992,Cirac:1992,Birkl1994,Monroe1995a,Morigi2000,Roos2000,Ejtemaee2016a} have been developed and employed in conjunction with Doppler cooling to meet the demands posed by different applications.

In experiments with trapped ions, sideband cooling on a narrow atomic transition \cite{Diedrich1989} or Raman sideband cooling \cite{Monroe1995a} on a broad transition with far off-resonant laser beams are both techniques capable of preparing an ion in the lowest vibrational state of the trapping potential. As these cooling techniques are applicable only to ions already prepared at low temperatures, they are performed after Doppler cooling \cite{Wineland1978,Neuhauser1978} the ions to temperatures where they are in the Lamb-Dicke regime, i.e. localized to much better than an optical wavelength.

Cooling by electromagnetically induced transparency (EIT) \cite{Morigi2000} is yet another ground-state cooling technique, which is applicable to pre-cooled ions. As with Raman sideband cooling, EIT requires an atomic three-level system coupled by a pair of laser beams. However, the laser detunings are chosen such that, to first order, the atomic system is pumped into a dark state, which does not couple to the laser fields. Because of this property, spontaneous photon emission rates are significantly lower than for Raman sideband cooling, and therefore, so are the heating processes intrinsic to the laser cooling scheme. Consequently, the range of oscillation frequencies over which the ion motion is cooled to close to the ground state can be made wider than with Raman sideband cooling, and higher cooling rates can be achieved. In comparison to sideband cooling on narrow transitions, EIT cooling also allows one to cool the ions over a wider frequency range to low energies and has the additional benefit of requiring far less laser power. 

EIT cooling was first experimentally demonstrated with a single trapped ion \cite{Roos2000}. More recently, EIT has been used to cool crystals of two and four ions \cite{Lin2013}, neutral atoms in a cavity-QED setup \cite{Kampschulte2014} and in a quantum-gas microscope \cite{Haller2015}. It has also been proposed as a technique for cooling nanomechanical oscillators \cite{Xia2009}.

In the context of quantum information processing and quantum simulation with ion strings, cooling one or several vibrational modes to close to the ground state is a prerequisite for most entanglement-generating atom-light interactions. Controlled-NOT gates with single-ion addressing \cite{Cirac1995} require a mode to be prepared in the ground state. The widely used entangling gates based on spin-dependent forces \cite{Sorensen2000} in principle only need the vibrational state to be prepared in the Lamb-Dicke regime, but experiments show that their performance improves \cite{Kirchmair2009} the lower the vibrational occupation number becomes. These applications make EIT cooling a powerful technique.

EIT cooling becomes even more appealing in the context of recent experiments simulating quantum Ising models in strings of up to eighteen ions \cite{Jurcevic2014,Richerme2014,Senko2014}. In these experiments all $2N$ radial modes of an $N$-ion string need to be cooled to close to the ground state. While sideband cooling can be used for cooling the modes sequentially, the task of multi-mode cooling is particularly well suited for EIT cooling. As we show in this paper, all the modes can be ground-state cooled with a single EIT cooling pulse in less than \unit[100]{$\mu s$} if they are bunched together in frequency space in a range of a few hundred kHz.

In this article, we give a description of our implementation of EIT cooling for trapped ions. The distinct broadband cooling capability, which is of considerable interest for all experiments having to deal with a dense mode spectrum, is demonstrated in a two-ion crystal in a segmented linear trap and in a nine and eighteen-ion crystal in a macroscopic linear trap. Sec.\:\ref{sec:theo} briefly summarizes the theory underlying EIT cooling. In Sec.\:\ref{sec:setup}, we introduce the two experimental setups used, detailing briefly the shared methods and procedures before separately presenting the features of each of the two apparatuses and describing the cooling experiments in Sec.\:\ref{sec:expt}. Sec.\:\ref{sec:conc} concludes the article with a discussion of the results. 

%%%%%%%%%%%%%%%%%%%%%%%%%%%%%%%%%%%%%%%%%%%%%%%%%%%%
\section{\label{sec:theo}EIT cooling mechanism}
	
We briefly review the theory of the EIT cooling based on ref. \cite{Morigi2000} and discuss its implementation with $^{40}\mathrm{Ca}^{+}$, the species used in the experiments presented.

\begin{figure}[t] \centering
\includegraphics[width = 1\linewidth]{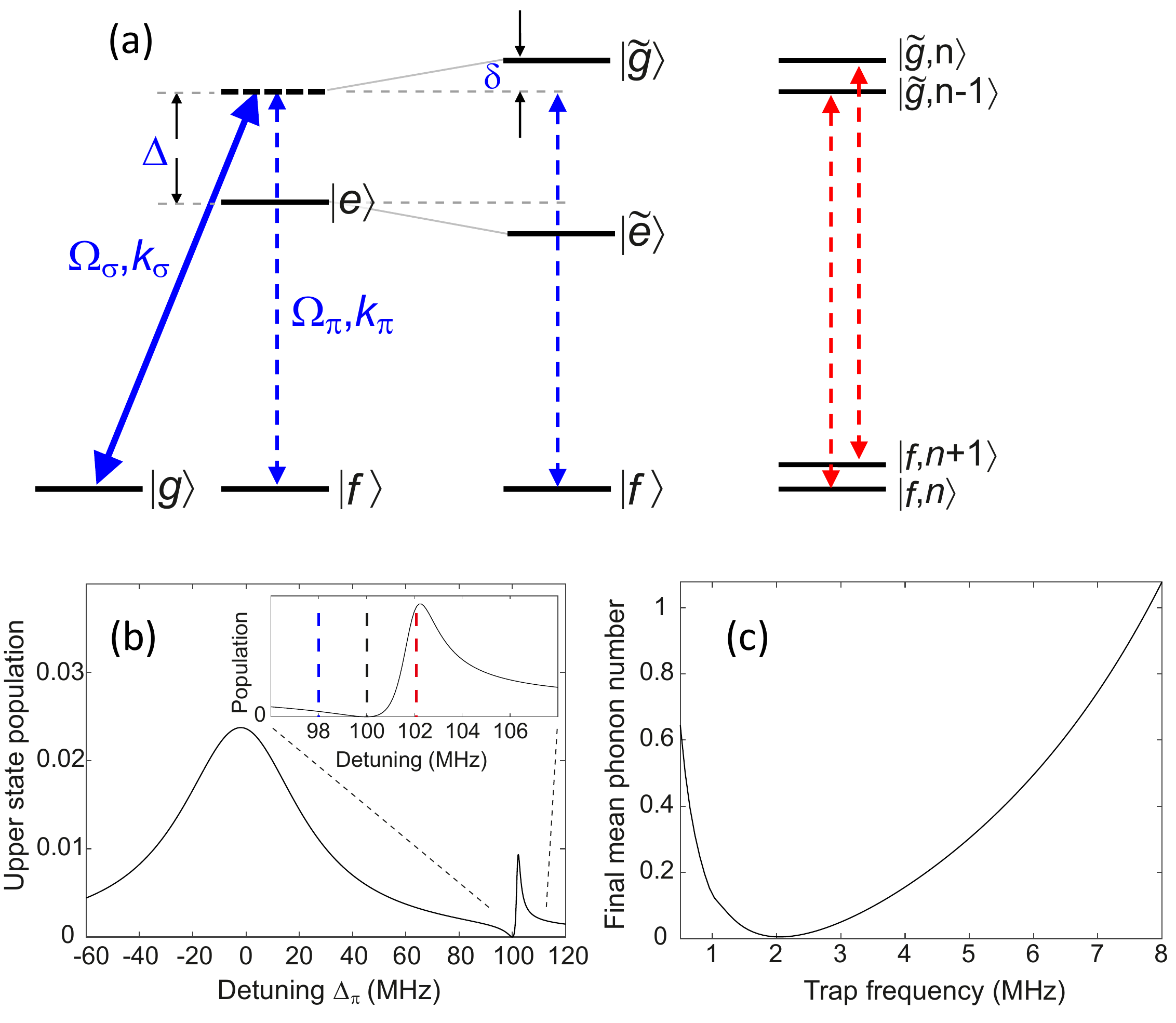}
\caption{\label{fig:levelscheme} (a) EIT cooling implemented in a three-level-$\Lambda$ system. One transition is driven by a strong, blue-detuned laser (Rabi frequency $\Omega_\sigma$) creating dressed states. A weaker laser probes the resulting Fano-like absorption profile (b) where the coupling to the dressed state $|\tilde{g}\rangle$ results in a narrow absorption feature. For a probe laser with a detuning $\Delta_\pi$ equal to the one of the dressing laser ($\Delta/(2\pi)=100$~MHz), absorption on the carrier transition is suppressed in steady state, and absorption on the red sideband $|f,n\rangle\Leftrightarrow |\tilde{g},n-1\rangle$ is much stronger than on the blue sideband (right-hand side of (a) and dashed lines in the inset of (b)). (c) Cooling to the ground state can be achieved over a wide range of trap frequencies as shown for the parameters in (b).} 
\end{figure}

EIT cooling requires a three-level $\Lambda$ atomic system where the coupling of two ground states, $|g\rangle$ and $|f\rangle$, to one short-lived excited state $|e\rangle$ leads to coherent population trapping in a superposition of the two ground states that does not couple to the excited state. Figure~\ref{fig:levelscheme}(a) shows the basic EIT scheme: the \mbox{$|g\rangle\Leftrightarrow |e\rangle$} transition is driven with a Rabi frequency, $\Omega_\sigma$, by a laser beam which is blue-detuned by an amount $\Delta$. The driving field creates dressed states $|\tilde{g}\rangle$, $|\tilde{e}\rangle$, which are light-shifted upwards or downwards in frequency by an amount
\begin{equation}
\delta = \frac{1}{2} \left(\sqrt{\Omega_{\sigma}^2 + \Delta^2}-\left| \Delta\right| \right) 
\label{eq:lightshift}
\end{equation}
from the bare states (where bare states are supposed to include the energy of the photons of the dressing beam), and pumps the atom to state $|f\rangle$. A probe beam drives the $|f\rangle\Leftrightarrow |e\rangle$ transition with detuning $\Delta_\pi$ and Rabi frequency $\Omega_\pi\ll\Omega_\sigma$. The coupling of the second ground state $|f\rangle$ to the dressed states creates a Fano-like absorption profile (Fig.~\ref{fig:levelscheme}(b)) \cite{Lounis1992}. For $\Delta_\pi=\Delta$, the absorption becomes zero when the atom is held at a fixed position in space. If, however, the atom is trapped in a harmonic trap with oscillation frequency $\omega$, it can absorb light on vibrational sidebands of the transition.
Setting the light shift $\delta$ of the dressing beam equal to the trap frequency, $\delta=\omega$, maximizes the absorption probability on the red sideband transition $|f,n\rangle\Leftrightarrow |\tilde{g},n-1\rangle$.  The phonon number $n$ decreases by one unit for every absorption event. As the absorption on the carrier transition $|f,n\rangle\Leftrightarrow |\tilde{g},n\rangle$ vanishes, the only remaining heating mechanism in the Lamb-Dicke regime is absorption on the blue sideband which can be made much smaller than red sideband absorption (see inset of Fig.~\ref{fig:levelscheme}(b)). 
%
%-------------------------------------------
%-------------------------------------------

The range of oscillation frequencies over which EIT cooling is efficient is set by choosing the detuning $\Delta$, which in turn determines the decay rate of the dressed state $|\tilde{g}\rangle$. Therefore, similar to other sideband cooling schemes, there is a trade-off between the lowest achievable temperatures and the range of oscillation frequencies over which cooling can be achieved. However, due to the suppressed carrier excitation, cooling close to the ground state is possible over a fairly large cooling range. The dynamics of the cooling process is described by a differential equation for the mean phonon number $\bar{n}$ \cite{Eschner2003},
\begin{equation}
 \dot{\bar{n}} = - \eta^{2} (A_{-}-A_{+})\bar{n} + \eta^{2} A_{+}.
\label{eq:phononnumber}
\end{equation}
\noindent Here, $\eta = \left|(\mathbf{k}_{\pi} -\mathbf{k}_{\sigma})\cdot\mathbf{e}_m \right| \sqrt{\frac{\hbar}{2 m \omega}}$ denotes the Lamb-Dicke factor with wave vector $\mathbf{k}_{\sigma}$ ($\mathbf{k}_{\pi}$) for the dressing (probe) laser, and $\mathbf{e}_m$ the unit vector describing the oscillation direction of the mode to be cooled. Furthermore, $m$ denotes the ion mass, and $\hbar=h/(2\pi)$ where $h$ is Planck's constant. The rate coefficients  $A_{\pm}$ are given by
\begin{equation}
A_{\pm} = \frac{\Omega_{\pi}^{2}}{\Gamma} \frac{\Gamma^{2} \omega^{2}}{\Gamma^{2} \omega^{2}+4\left(\frac{\Omega_{\sigma}^{2}}{4}-\omega(\omega \mp \Delta)\right)^{2}}, 
%\label{eq:coeffs}
\end{equation}
where $\Gamma$ is the linewidth of the transition \cite{Morigi2000}. The steady-state solution of Eq.~(\ref{eq:phononnumber}) is $\left\langle n\right\rangle= \frac{A_{+}}{A_{-}-A_{+}}$ and the cooling rate is given by $R=\eta^2(A_--A_+)$. An example of the achievable cooling range is shown in Fig.~\ref{fig:levelscheme}(c).

 In $^{40}\mathrm{Ca}^{+}$, a good approximation to such a three-level system is realized by using the Zeeman sublevels of the $\mathrm{S}_{1/2} \Leftrightarrow \mathrm{P}_{1/2}$ dipole-transition at \unit[397]{nm} \cite{Roos2000} for EIT cooling, as illustrated in Fig.\:\ref{fig:align}(a). Dressed states are generated by \mbox{$\sigma^{+}$-polarized} light strongly coupling the $\left|\mathrm{S}_{1/2},m=-1/2\right\rangle$ ground state to the excited state $\left|\mathrm{P}_{1/2},m=1/2\right\rangle$. 
A \mbox{$\pi$-polarized} laser beam coupling the $\left|\mathrm{S}_{1/2},m=+1/2\right\rangle$ state to the $\left|\mathrm{P}_{1/2},m=+1/2\right\rangle$ probes the absorption spectrum modified by the dressing laser.
The detunings of the EIT dressing and probe beams have to be adjusted such that their difference matches the Zeeman splitting of the ground state in the presence of a magnetic field. The \mbox{$\pi$-polarized} probe beam also off-resonantly couples the states $\left|\mathrm{S}_{1/2},m=-1/2\right\rangle$ and $\left|\mathrm{P}_{1/2},m=-1/2\right\rangle$. However, the additional heating introduced by this coupling is negligible compared to other heating processes of the EIT scheme as there is little population in $\left|\mathrm{S}_{1/2},m=-1/2\right\rangle$ due to optical pumping by the dressing laser. Of more serious concern are polarization imperfections of the EIT beams, which can depump the $\left|\mathrm{P}_{1/2},m=-1/2\right\rangle$ state and thus give rise to heating. Such polarization imperfections may arise due to geometric constraints in the optical access to the ion, as is the case in the experiments described in this article.

With a linewidth $\Gamma_\mathrm{PS}/(2\pi)=\unit[21.57(8)]{MHz}$ \cite{Hettrich2015}, the $\mathrm{S}_{1/2} \Leftrightarrow \mathrm{P}_{1/2}$ transition enables fast cooling close to the ground state as was shown for a single ion in ref.\cite{Roos2000}.

%%%%%%%%%%%%%%%%%%%%%%%%%%%%%%%%%%%%%%%%%%%%%%%%%%%%
\section{\label{sec:setup}Experimental setups}

For the experiments presented here, two different kinds of  radio-frequency (rf) trap design were used: a macroscopic linear trap (setup 1) and a micro-fabricated segmented two-layer trap (setup 2). In this section, the two setups are described, followed by a presentation of the procedures used to calibrate the parameters of the EIT cooling laser beams in both experiments.

\begin{figure}[t] \centering
\includegraphics[width = 0.8\linewidth]{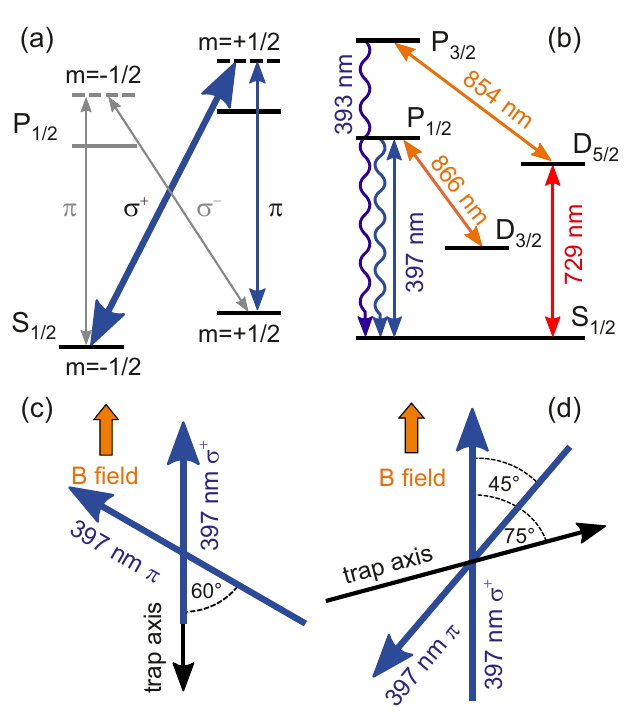}
\caption{\label{fig:align} (a-b) Schematic of the $^{40}\mathrm{Ca}^{+}$ transitions used in the experiment. EIT cooling is implemented using the Zeeman sublevels of the $\mathrm{S}_{1/2} \Leftrightarrow \mathrm{P}_{1/2}$ dipole transition. The dipole-allowed transitions for Doppler cooling, detection and quenching are excited by light at \unit[397]{nm}, \unit[866]{nm} and \unit[854]{nm} while the quadrupole transition that is used for motional state analysis is driven by light at \unit[729]{nm}. Beam configuration for EIT cooling in setup 1 (c) and setup 2 (d). The trap axis denotes the direction in which the rf-zero field line is oriented.} 
\end{figure}

The releveant energy levels of $^{40}\mathrm{Ca}^{+}$ are sketched in Fig.\:\ref{fig:align}\:(b). Ions are loaded into the trap by two-step photoionization of a beam of neutral calcium atoms using lasers operating at \unit[422]{nm} and \unit[375]{nm}. A laser at \unit[397]{nm} Doppler-cools trapped ions on the $\mathrm{S}_{1/2} \Leftrightarrow \mathrm{P}_{1/2}$ transition. A laser at \unit[866]{nm} is used to repump population from the metastable $\mathrm{D}_{3/2}$ state. The narrowline quadrupole transition $\mathrm{S}_{1/2} \Leftrightarrow \mathrm{D}_{5/2}$ is used for temperature diagnostics. It is driven by a Ti:Sa laser emitting at \unit[729]{nm}, which is stabilized to a high-finesse resonator with a finesse of a few hundred thousand, in order to achieve a laser linewidth $<\unit[10]{Hz}$. The $\mathrm{D}_{5/2}$ state's lifetime of about \unit[1.1]{s} \cite{Barton2000} is quenched by applying laser light at \unit[854]{nm} which couples the metastable state to the short-lived $\mathrm{P}_{3/2}$ state. Read-out of the electronic state of the ions is carried out using the electron-shelving method \cite{Bergquist1986,Nagourney1986,Sauter1986} with a photo-multiplier tube or an electron-multiplying charged coupled device (EMCCD) camera for a measurement of the total or individual states of the ions.

The EIT cooling beams are derived from the same laser that is used for Doppler-cooling the ions. While the geometries of the two experimental setups give rise to different beam configurations for the $\sigma^{+}$ and the $\pi$ beam, as detailed below, the optical setups share the same principle: A part of the available \unit[397]{nm} light is split into two separate beam paths. Two independently programmable acousto-optic modulators create laser pulses of precisely controlled frequency, intensity and duration, which are subsequently sent through single-mode fibers to the vacuum chamber where they illuminate the ion crystal.

%-----------------------------
%-----------------------------

%%%%%%%%%%%%%%%%%%%%%%%%%%%%%%%%%%%%%%%%%%%%%%%%%%%%
\subsection{\label{ssub:lintrap}Setup 1: macroscopic linear trap}
	
EIT cooling of long strings of ions was investigated in a macroscopic linear ion trap with an endcap-to-endcap electrode distance of \unit[4.5]{mm} and a minimum electrode-ion separation of \unit[565]{\textmu m} \cite{Hempel2014}. Two coils connected to a stabilized-current drive generated a magnetic field of \unit[0.4]{mT}, oriented along the rf-zero field line of the trap. The geometry of the magnetic field direction and the laser beam $k$-vectors are shown in Fig.\:\ref{fig:align}\:(c). The frequency difference of the EIT cooling beams was set to match the resulting Zeeman ground state splitting of  \unit[11.4]{MHz}. The \mbox{$\sigma^{+}$-polarized} EIT dressing beam was detuned from the $\mathrm{S}_{1/2} \Leftrightarrow \mathrm{P}_{1/2}$ Zeeman transition by \unit[106]{MHz}. The wave vector of the EIT probe beam propagated at an angle of \unit[60]{\degree} to the magnetic-field vector. The Raman $k$-vector of the EIT beams (set by their $k$-vector difference) formed an angle of \unit[60]{\degree} with the axial direction of the trap and had equal overlap with both principal axes of the radial trapping potential. The EIT probe beam was not perfectly \mbox{$\pi$-polarized} because it was not normal to the quantization axis. 

%%%%%%%%%%%%%%%%%%%%%%%%%%%%%%%%%%%%
\subsection{\label{ssub:segtrap} Setup 2: segmented trap}

A two-layer gold-on-alumina sandwich trap \cite{Harlander2012} with an ion-electrode separation of \unit[258]{\textmu m} and a segment length of \unit[250]{\textmu m} was used to trap a crystal of two $^{40}\mathrm{Ca}^{+}$ ions. Four coils connected to stabilized-current drives generated a magnetic field of \unit[0.33]{mT} aligned at an angle of \unit[75]{\degree} to the trap axis (see Fig.\:\ref{fig:align}\:(d)). The magnetic field lifted the degeneracy of the ground Zeeman levels, resulting in a level splitting of \unit[9.3]{MHz}. 

The frequency difference of the two EIT cooling beams was chosen to match the ground-state state Zeeman splitting. The detuning of the \mbox{$\sigma^{+}$-polarized} EIT beam from the \mbox{$\mathrm{S}_{1/2} \Leftrightarrow \mathrm{P}_{1/2}$} Zeeman transition that it was driving was \mbox{$\Delta/(2\pi) \approx \unit[105]{MHz}$}. The $\sigma^{+}$ beam was aligned parallel to the magnetic field. The $k$-vector of the \mbox{$\pi$-polarized} light had an angle of \unit[45]{\degree} to the magnetic field. This choice was dictated by constraints in the available optical access and led to imperfect $\pi$ polarization, with the additional polarization components heating the ion. The trap axis lay in the plane spanned by the two EIT cooling beams. It formed an angle of \unit[75]{\degree} with the direction of the magnetic field and an angle of \unit[52.5]{\degree} with the Raman $k$-vector of the EIT beams. The two principal radial directions (mutually orthogonal with weak axis) both had an overlap with the Raman $k$-vector. All three sets of vibrational modes were therefore accessible to EIT-cooled in this geometry. 

A segmented trap like the one used in setup 2 is in principle capable of creating anharmonic axial potentials for the creation of ion strings with approximately equidistantly spaced ions \cite{Lin2009}. Technical difficulties with the particular trap made it difficult to reliably store long ion crystals; for this reason, the majority of the EIT cooling experiments were carried out in setup 1.
	
%%%%%%%%%%%%%%%%%%%%%%%%%%%%%%%%%%%%
\subsection{\label{ssub:calibration}EIT beam parameter calibration}

For EIT cooling, the frequency, intensity and polarization of both cooling beams have to be set properly. In the following, we will discuss different methods for measuring the laser beam parameters.

\emph{Frequency calibration.---} We set the frequency difference of the beams by measuring the ground-state Zeeman splitting  spectroscopically on the $\mathrm{S}_{1/2} \Leftrightarrow\mathrm{D}_{5/2}$ quadrupole transition. The overall detuning of the beams from the $\mathrm{S}_{1/2} \Leftrightarrow\mathrm{P}_{1/2}$ dipole transition was set by measuring the spectral line profile of the $\mathrm{S}_{1/2}\Leftrightarrow \mathrm{P}_{1/2}$ transition with the Doppler cooling laser and frequency-shifting the EIT beams by the desired amount by acousto-optical modulators. As the EIT cooling beams were derived from the same laser as the Doppler cooling beam, the uncertainty in the determination of the EIT beam detuning was due to systematic errors in the line center measurement, which we estimated in our experiment to be below \unit[10]{MHz}.
 
\emph{Polarization calibration.---} The polarization of the EIT probe beam was set by aligning the beam's polarization at the entrance of the vacuum system with the magnetic field by rotating the polarization into the plane spanned by the magnetic field vector and the beam's $k$-vector.

The polarization of the EIT dressing beam was controlled by a quarter-wave plate in front of the viewport, converting the linear polarization coming out of an optical fibre into $\sigma^+$-polarization. In setup 2, the polarization was adjusted by minimizing the population in the $\left|\mathrm{S}_{1/2},m=-1/2\right\rangle$ state, detected by carrying out a $\pi$ pulse on the $\left|\mathrm{S}_{1/2},m=-1/2\right\rangle \Leftrightarrow \left|\mathrm{D}_{5/2},m=-1/2\right\rangle$ transition followed by a measurement of the $\mathrm{D}_{5/2}$ population. We judged this measurement to be more precise than a minimization of the fluorescence signal induced by the $\sigma^{+}$ beam. 
An even better technique, applied in setup 1, wis discussed at the end of this section.

\emph{Intensity calibration.---} Optimal EIT cooling of a vibrational mode of frequency $\omega$ requires making the light shift $\delta$, induced by the \mbox{$\sigma^{+}$-polarized} light, equal to $\omega$. For a fixed detuning $\Delta$ of the EIT beams, the light shift was set by controlling the intensity ($\propto\Omega_\sigma^2$) of the dressing beam (see Eq.~(\ref{eq:lightshift})). Two different intensity calibration techniques were employed in the experiments. First, light shifts can be precisely determined by measuring the optical pumping rate between $\mathrm{S}_{1/2}$ Zeeman states \cite{Hettrich2015}. This technique (used in setup 2) relies on a precise knowledge of the laser parameters and the excited state's decay rate. 

%-----------------------------
\begin{figure}[t] \centering
\includegraphics[width = 1\linewidth]{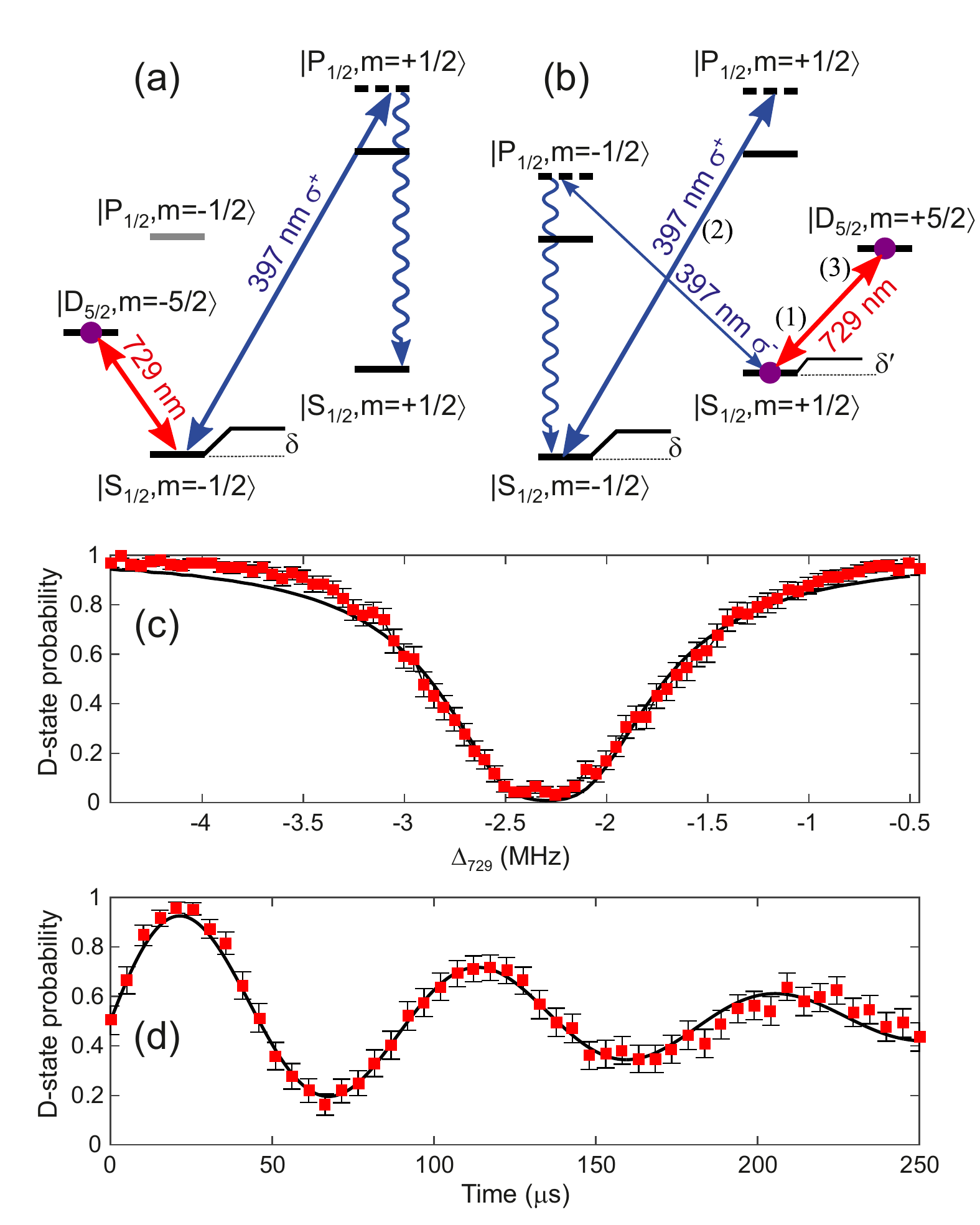}
%
% DATA: c: Light shift measurement: 20141126-1055 (18 ions, LightShiftDetermination_Data_plusSimulation.m)
% DATA: d: polarization optimization: 20141203-1340 (9 ions, FourLevelAtom_timeevol_RamseyExp_plusData.m)
\caption{\label{fig:lightshift} (a) Measurement of the EIT light shift. After preparing the ion in $\left|\mathrm{D}_{5/2},m=-5/2\right\rangle$, the EIT dressing beam and the 729 nm laser are simultaneously switched on for a duration $\tau$. The light shift induced by the dressing laser on the $\left|\mathrm{S}_{1/2},m=-1/2\right\rangle$ state is probed by measuring the D-state population as a function of the detuning $\Delta_{729}$ from the Zeeman transition probed by the laser. Measured data is shown in (c). The light shift is equal to the detuning, $\Delta_{729}$, for which the $\mathrm{D}_{5/2}$-state population is maximally depleted. (b) Optimization of the EIT dressing beam polarization. Unwanted $\pi$- and $\sigma^-$- polarization components will frequency-shift the $\left|\mathrm{S}_{1/2},m=1/2\right\rangle$ state by an amount $\delta^\prime$ and pump it out. These effects can be probed by sandwiching the EIT dressing pulse into a Ramsey experiment on the $\left|\mathrm{S}_{1/2},m=-1/2\right\rangle\Leftrightarrow \left|\mathrm{D}_{5/2},m=5/2\right\rangle$ transition. (d) The $\mathrm{D}_{5/2}$-state population measured as a function of the dressing pulse length oscillates at frequency $\delta^\prime$ and thus provides information about the strength of the residual polarization components. In this example, the intensity of the residual components was 200 times weaker than the intensity of the \mbox{$\sigma^+$-polarized} component.} 
\end{figure}
%-----------------------------

Another, more direct way, of measuring the light shift exploits spectroscopy on the $\mathrm{S}_{1/2} \Leftrightarrow\mathrm{D}_{5/2}$ transition. It uses the fact that the \mbox{$\sigma^{+}$-polarized} blue-detuned EIT-dressing-beam frequency-shifts the states $\left|\mathrm{P}_{1/2},m=+1/2\right\rangle$ and $\left|\mathrm{S}_{1/2},m=-1/2\right\rangle$ by the same amount but in opposite directions, for detunings much smaller than the fine-structure splitting. This statement approximately holds even for arbitrary polarization if the detuning is large compared to the Zeeman splitting. For a measurement of the $\left|\mathrm{S}_{1/2},m=-1/2\right\rangle$ state's light shift, the ion is initially prepared in the $\left|\mathrm{D}_{5/2},m=-5/2\right\rangle$ state (Fig.~\ref{fig:lightshift}(a)). Next, the EIT dressing beam is switched on simultaneously with a laser pulse at \unit[729]{nm} driving the $\left|\mathrm{S}_{1/2}, m=-1/2\right\rangle \Leftrightarrow \left|\mathrm{D}_{5/2},m=-5/2 \right\rangle$ transition. This was followed by a measurement of the D-state probability as a function of the \unit[729]{nm} laser frequency. As this scheme depletes the $ \left|\mathrm{D}_{5/2},m=-5/2 \right\rangle$ state by optically pumping the population into $\left|\mathrm{S}_{1/2}, m=+1/2\right\rangle$, the light shift was determined by measuring the shift of the \unit[729]{nm} laser frequency from the unperturbed transition frequency at which the probability for being in the $D_{5/2}$ state is at a minimum. 

Figure~\ref{fig:lightshift}(c) shows results from an application of the technique (in setup 1) for the case of a dressing laser inducing a shift of about \unit[2.3]{MHz}. For a 729 nm laser pulse with Rabi frequency $\Omega=(2\pi)\,\unit[39]{kHz}$, a pulse duration of \mbox{$\tau\,$=\,\unit[250]{\textmu s}} was found to optimally deplete the $\mathrm{D}_{5/2}$ state without causing unnecessary broadening of the resonance. The measured excitation was fitted by numerically simulating the dynamics using a master-equation approach. The same calibration method was also employed to set the intensity of the EIT probe beam to a value much smaller than the dressing-beam intensity.

A variation of this technique can be used for optimizing the $\sigma^{+}$ polarization of the EIT dressing beam as is illustrated in Fig.~\ref{fig:lightshift}(b),(d). A Ramsey experiment was carried out on the $|\mathrm{S}_{1/2},m=1/2\rangle\Leftrightarrow |\mathrm{D}_{5/2},m=5/2\rangle$ transition with a \unit[397]{nm} EIT dressing pulse of variable duration sandwiched between the two Ramsey $\pi/2$ pulses with optical phases shifted by $\pi/2$ with respect to each other. If the blue light is perfectly  \mbox{$\sigma^{+}$-polarized}, it will have no effect on the superposition state, resulting in a final $D_{5/2}$ state population of 0.5 irrespective of the pulse duration. If, however, the pulse is imperfectly polarized, then  the pumping and light-shifting action of the undesirable $\sigma^{-}$ or $\pi$ polarization components will give rise to an exponentially decaying oscillatory signal as a function of the length of the EIT dressing pulse. This technique enables a very precise suppression of spurious polarization components as the data shown Fig.\:\ref{fig:lightshift}\:(d) demonstrates. In this experiment, which used nine ions, optimization of the polarization reduced unwanted light shifts by wrong polarization components to \unit[10.8(5)]{kHz} whereas the desired polarization component induced a shift of about \unit[2200]{kHz}, which was inferred from an independent experiment as depicted in Fig.~\ref{fig:lightshift}(a),(c). The measurement shows that the technique enables the detection of unwanted polarization components with fractional power well below $10^{-3}$.
		
%%%%%%%%%%%%%%%%%%%%%%%%%%%%%%%%%%%%%%%%%%%%%%%%%%%%
\section{\label{sec:expt}Experimental results}
%-----------------------------
\begin{figure*}[htb] \centering
\includegraphics[width = \linewidth]{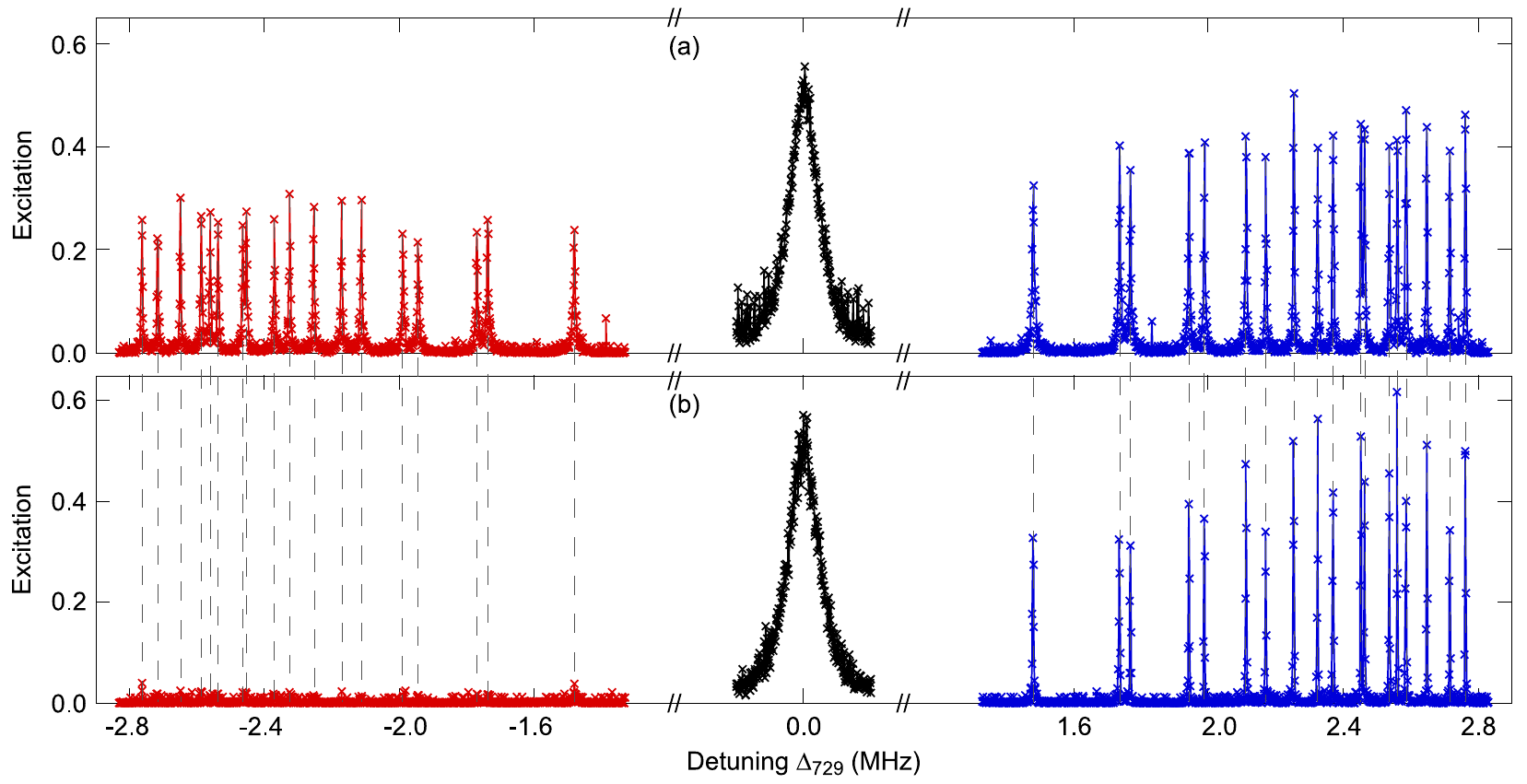}
%
% DATA: 9-ion spectra: 20141202 (PlotSpectra_9_ions_v2.m or PlotSpectra9ion_20140212V6)
\caption{\label{fig:data9spec} (a) Measured radial mode spectrum of a nine-ion linear crystal after Doppler cooling. (b) Radial mode spectrum of a nine-ion linear crystal after Doppler cooling and subsequent EIT cooling of \unit[1]{ms} duration with a $\sigma^{+}$ induced light shift set to \unit[2.2]{MHz}. Data points are connected by solid lines to enhance the visibility of the resonances. Dashed lines connecting (a) and (b) indicate the mode frequencies. The disappearance of all red sidebands demonstrates that the EIT pulse simultaneously cools all radial modes, which are spread over a range of \unit[1.2]{MHz}. The weak excitation in the wings of the carrier transition visible in (a) but not in (b) are most likely due to second-order sideband processes exchanging phonons between two radial modes.} 
\end{figure*}
%-----------------------------
The basic experimental sequence for investigating EIT cooling consists of the following steps: (i) the ions are Doppler-cooled on the $\mathrm{S}_{1/2}\Leftrightarrow\mathrm{P}_{1/2}$ transition for a few milliseconds. (ii) The \unit[397]{nm} EIT cooling beams are applied simultaneously with the laser at \unit[866]{nm} serving as repumper. (iii) A \unit[50]{\textmu s} optical pumping pulse with \mbox{$\sigma^{+}$-polarized} light and light at \unit[866]{nm} ensure that the ion is prepared in the $\left|\mathrm{S}_{1/2},m=1/2\right\rangle$ state. (iv) The state of a single motional mode is analyzed by excitation of the corresponding first vibrational sideband of the $\mathrm{S}_{1/2}\Leftrightarrow\mathrm{D}_{5/2}$ quadrupole transition, followed by a measurement of the electronic states of the ions.

%%%%%%%%%%%%%%%%%%%%%%%%%%%%%%%%%%%%
\subsection{\label{ssub:linres} EIT cooling of multi-ion crystals}

We investigated EIT cooling of linear ion crystals held in strongly anisotropic harmonic potentials, which result in a radial mode spectrum whose width is considerably smaller than its mean mode frequency. This configuration is of interest in experiments where a laser-induced coupling of the radial modes to the internal states of the ions is used either for studying spin-boson models or long-range spin-spin interactions mediated by the radial modes \cite{Schneider2012,Richerme2014,Jurcevic2014}. A distinctive feature of this configuration is the low axial center-of-mass (COM) frequency, which is typically about an order of magnitude smaller than the radial center-of-mass frequencies.

In setup 1, measurements were taken using a crystal of nine ions with COM mode frequencies of \unit[$\left\{\omega_{z},\omega_{r1},\omega_{r2}\right\}=2 \pi \left\{0.50, 2.59, 2.76\right\} $]{MHz} and a crystal of eighteen ions with COM modes frequencies of \unit[$\left\{\omega_{z},\omega_{r1},\omega_{r2}\right\}=2 \pi \left\{0.21, 2.68, 2.71\right\}$]{MHz}. Note that the COM modes are the highest-frequency radial modes, but the lowest-frequency axial mode.

For the nine-ion crystal, all radial modes were cooled simultaneously by centering the light shift induced by the dressing beam roughly in the middle of the radial mode spectrum at \unit[2.2]{MHz}, which corresponded to setting \mbox{$\Omega_\sigma=(2\pi)\,$\unit[30]{MHz}}. The intensity of the probe beam was set such that $\Omega_\pi=(2\pi)\,$\unit[6.2]{MHz} $\ll\Omega_\sigma$.

Figure~\ref{fig:data9spec} compares the radial sideband spectrum of a Doppler-cooled nine-ion crystal with a spectrum obtained after an additional EIT cooling pulse of \unit[1]{ms} duration. The complete vanishing of the red sideband absorption in case of the EIT-cooled ion crystal indicates that all 18 radial modes were simultaneously cooled to the ground state of motion. 

\begin{figure}[t] \centering
\includegraphics[width =  \linewidth]{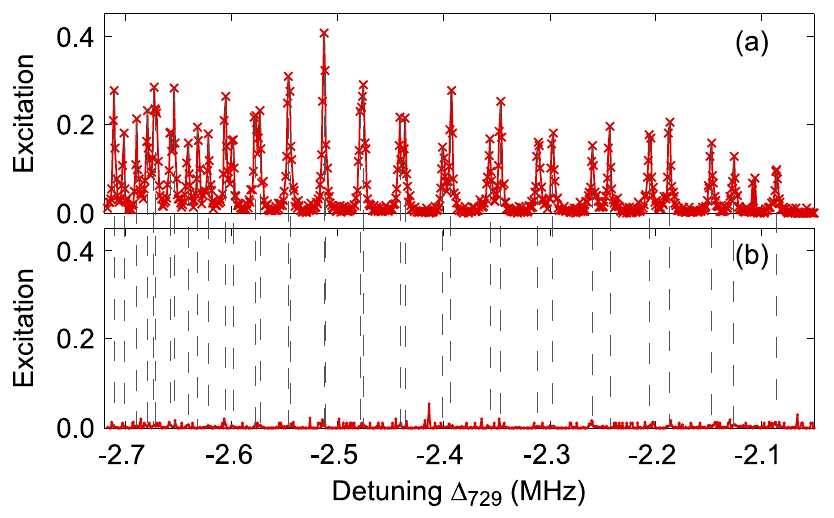}
%
% DATA: 18-ion spectrum: 20141126 (PlotSpectra_18_ions.m or PlotSpectra18ion_20141126V3.m)
\caption{\label{fig:data18spec} (a) Red-sideband radial mode spectrum of an eighteen-ion crystal after Doppler cooling. (b) Red-sideband radial mode spectrum after Doppler cooling followed by \unit[1]{ms} of EIT cooling with a $\sigma^{+}$ induced light shift set to \unit[2.3]{MHz}. Dashed lines connecting (a) and (b) indicate the mode frequencies.} 
\end{figure}

 EIT cooling of all radial modes was also observed for an eighteen-ion crystal. Figure~\ref{fig:data18spec} shows the radial red sideband absorption spectrum after Doppler cooling (upper plot) and after an additional \unit[1]{ms} EIT cooling pulse (lower plot). Again, absorption on all 36 radial sidebands completely disappears.

The strong reduction of the red-sideband absorption is indicative of ground-state cooling. However, the comparison of the absorption probabilities on the red sideband, $p_r$, and the blue sideband, $p_b$, of a vibrational mode does not yield a simple quantitative estimate of the mean phonon number. This is in contrast to the case of a laser addressing a single ion only, where $\bar{n}=p_r/(p_b-p_r)$. This formula is no longer valid if all ions are excited simultaneously as can be seen from the Doppler-cooled sideband spectrum of Fig.\:\ref{fig:data9spec}\:(a), where a marked absorption asymmetry between red and blue sidebands occurs even though Doppler cooling should lead to mean phonon number of about $\bar{n}\approx 5$, i.e. to a sthate where the coupling strengths on the red and blue sidebands are very similar for the majority of the phonon distribution.
%-----------------------------
%-----------------------------

In order to obtain an estimate of $\bar{n}$ after EIT cooling, we numerically simulated the red-sideband absorption spectrum of a nine-ion crystal for the parameters of our experiment. As the relevant state space of all vibrational modes was too big for numerical simulations ($\sim 10^{20}$), we calculated the spectrum in the vicinity of each mode by assuming that a spatially homogeneous laser beam coupled $N$ two-level atoms only to the vibrational mode of interest. This approach reduced the state space to sets of at most $N+1$ coupled states, $|\psi_k^n\rangle = (\hat{S}^+)^k|0,n\rangle$, where $\hat{S}^+$ is an electronic raising operator defined via $\hat{S}^+=\sum_{i=1}^N\eta_{im}\sigma_i^+a$ with $\eta_{im}$ the Lamb-Dicke factor of ion $i$ and mode $m$, and $|j,n\rangle$ denotes an $N$-ion state with $j$ electronic and $n$ vibrational excitations. When starting with all ions in the electronic ground state, the simulated spectra closely resembled the experimental results of Fig.~\ref{fig:data9spec}\:(a) for thermal phonon distributions with $\bar{n}\approx 5-8$. The blue-sideband spectra were simulated with a similar approach and also matched the observations. 

While such numerical simulations provide estimates of mean phonon numbers, it is hard to quantify the accuracy of the results. Alternatively, addressing of single ions in ion strings could be used to experimentally investigate motional states in multi-ion systems. However, apart from the technical difficulty of achieving single-ion addressing, such an approach would yield only one bit of information per measurement. 

For these reasons, we developed a novel experimental technique applicable to an $N$-ion crystal that yields single-shot measurements of the lowest Fock states (and provides $N$ bits of information): The cooling limit and cooling dynamics of the EIT method are investigated by means of rapid adiabatic passages (RAPs)  \cite{Lechner2016a}. In trapped-ion experiments, RAPs have been used for electronic-state manipulation \cite{Wunderlich2007,Toyoda2011} and for Fock-state measurements with a single ion \cite{An2015} by multiple RAPs.  For an ion string, a RAP on a red-sideband transition enables conversion of phononic into electronic excitations if the laser pulse is globally applied to all the ions in a crystal that is initially prepared in the electronic ground state. For the radial center-of-mass mode, the mapping is one-to-one for phonon numbers smaller than or equal to $N$. For modes with unequal Lamb-Dicke factors, the mapping works reliably for phonon numbers $\le N/2$ \cite{Lechner2016a}. Using this technique, we obtain single-shot measurements of the phonon number by detecting the number of electronically excited ions after the RAP. 

In the RAPs used in our experiments, we typically swept the laser frequency within \unit[4-5]{ms} over a frequency range of \unit[50]{kHz} centered around the sideband frequency. We applied a Gauss-like temporal intensity profile with a peak Rabi frequency of about \unit[200]{kHz} on the carrier transition. 

As an imperfect mapping of phononic to electronic excitations would make the cooling results look  better than they actually are, the validity of the mapping technique was checked by carrying out RAPs on the red sideband for an ion string prepared in the electronic {\it metastable} state instead of the electronic ground state. In this case, the RAP always couples sets of $N$+1 states together and therefore one expects to be able to perfectly invert the electronic state irrespective of the number of phonons. In experiments with nine ions, we indeed observed a near-perfect transfer of the ions to the electronic ground state when exciting the ions on the red sideband of a radial COM mode: in more than 80\% of the experiments, the state transfer was perfect; on average, 8.78(3) instead of 9 excitations were transferred.
% DATA: DataAnalysis_9_ions_heatingrate_20141001_plotresults.m

\begin{figure}[t] \centering
\includegraphics[width = \linewidth]{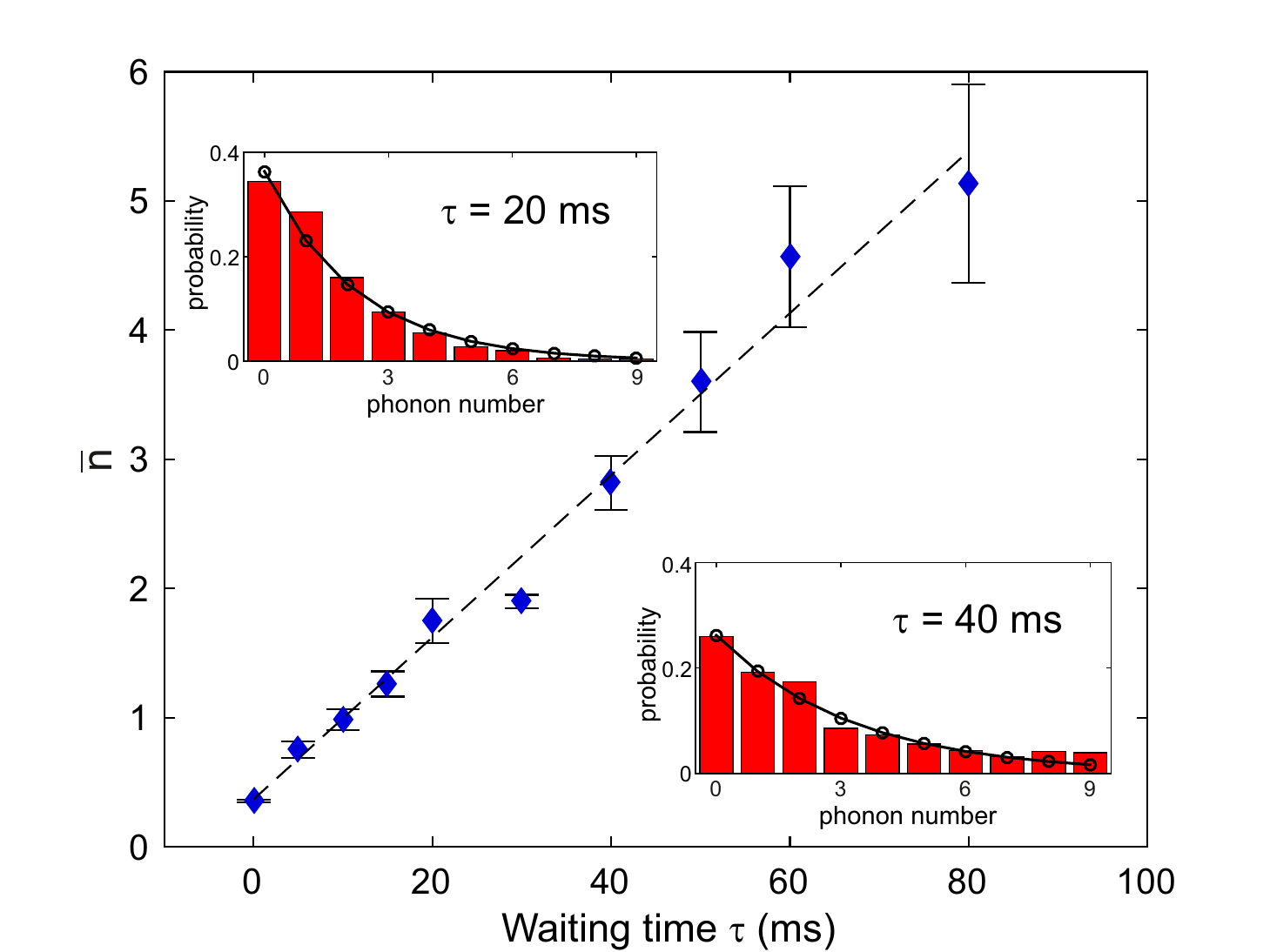}
%
% DATA: heating rate: 20141001, (9 ions, DataAnalysis_9_ions_heatingrate_20141001_plotresults.m)
\caption{\label{fig:data9heat} Heating rate measurement by sideband RAPs. After EIT cooling a nine-ion crystal, the mean phonon number after varying waiting times was measured using a RAP on the red sideband of a COM mode at \unit[2.74]{MHz}. The histograms show the measured phonon distributions after \unit[20]{ms} and \unit[40]{ms} of waiting time together with a fitted thermal distribution (solid line). A linear fit to the data (dashed line) yields a heating rate of \unit[$(65 \pm 7)$]{$\mathrm{s}^{-1}$}, which is about nine times higher than the heating rate observed for a single ion.}
\end{figure}

We determined the mean phonon numbers of various radial modes of EIT-cooled nine- and eighteen-ion crystals. In all cases we found mean phonon numbers of about 0.01 to 0.02 after EIT cooling. We also measured the enviroment-induced heating rate of the highest-frequency radial COM mode of a nine-ion crystal by introducing a variable delay between the EIT cooling and the RAP analysis. The data displayed in Fig.\:\ref{fig:data9heat} show a linear increase of the mean phonon number to about 5 after 80~ms wait time, corresponding to a heating rate of \unit[$(65 \pm 7)$]{$\mathrm{s}^{-1}$}. As the RAP technique enables single-shot phonon-number measurements, the data also confirms that the phonon distribution is thermal, as expected for a process coupling a ground-state cooled mode to a high-temperature reservoir. 

%-----------------------------
%-----------------------------

The measured nine-ion COM mode heating rate was about an order of magnitude higher than the measured heating rate of a single ion in the same trapping potential. This observation is consistent with our expectation to find a heating rate of the COM mode that scales linearly with the number of ions, as electric-field noise acting on different ions is nearly perfectly correlated in a macroscopic linear trap. Under these condition, all the energy taken up by the ions is deposited in the COM mode, thus heating this mode \cite{Home2013,Sawyer2014} at a much higher rate than the one, when there is only a single ion in the trap. Unexpectedly, higher-order radial modes showed a non-zero heating rate which seemed to become higher for shorter-wavelength modes. The origin of this source of heating is currently not clear; one possible source  might be a mode cross-coupling with the axial normal modes \cite{Marquet2003}, some of which are populated with tens or hundreds of phonons. We plan to investigate this issue in future experiments.

\begin{figure}[t] \centering
\includegraphics[width =\linewidth]{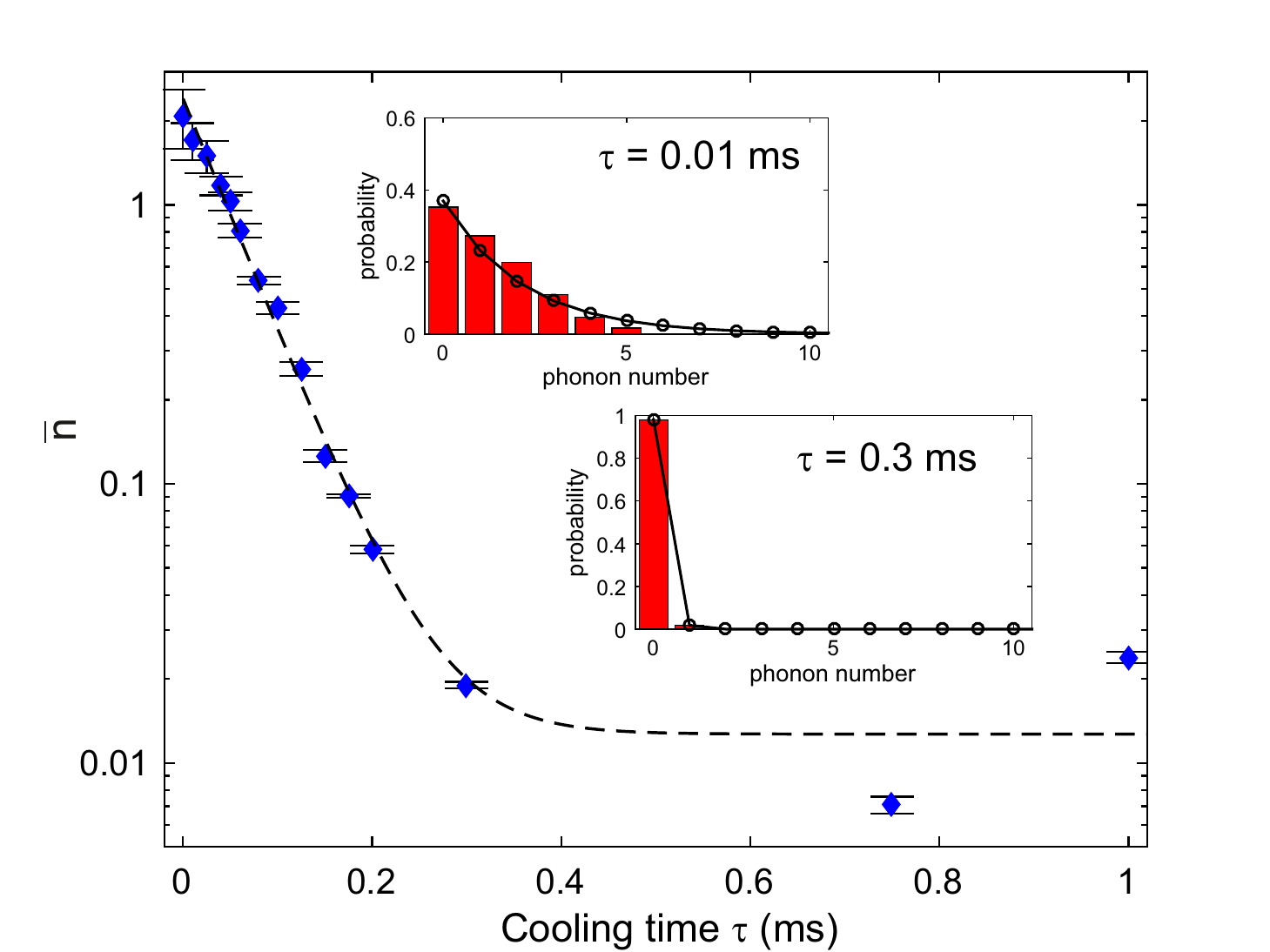}
%
% DATA: cooling rate: 20141126, (18 ions, CoolingDynamics18ions_20141126.m)
\caption{\label{fig:data18dyn}  EIT cooling dynamics. The radial mode at \unit[2.08]{MHz} of an eighteen-ion crystal in a linear trap was EIT cooled with a $\sigma^{+}$ induced light shift of \unit[2.3]{MHz} for varying cooling duration. A RAP pulse mapped the phonon number to the electronic state of the ions, which was subsequently measured. The histograms give the probability for observing a given number of excited ions running from none to all the ions within the crystal, as depicted for various cooling times in the insets. The mean phonon number was derived from a thermal fit to the data (solid line). The dashed line is a fit of the measured phonon numbers to the model described in the main text. It yields a cooling rate of \unit[$(19 \pm 3)10^{3}$]{$\mathrm{s}^{-1}$}.} 
\end{figure}

The EIT cooling dynamics was investigated with an eighteen-ion crystal: Doppler-cooled ions were subjected to an EIT cooling pulse of variable duration followed by a RAP measuring the resulting phonon distribution. Figure\:\ref{fig:data18dyn} shows the results obtained for the lowest-frequency radial mode at \unit[2.08]{MHz}. This mode of the ion string was cooled below a mean phonon number of one after less than \unit[50]{\textmu s} of EIT cooling, reaching a steady state after approximately \unit[300]{\textmu s}. For a fit of the dynamics, we assume that the time evolution of the mean phonon number $\bar{n}$ is given by $\dot{\bar{n}}=-R\bar{n}+R_{\mathrm{h}}$ where $R$ is the laser cooling rate and $R_h$ accounts for heating processes, resulting in $\bar{n}(t)=Ae^{-Rt}+n_{\mathrm{eq}}$. To extract the cooling rate, a least-square fitting routine was applied to the logarithm of the measured mean phonon number in order to properly weight the later points in the cooling dynamics. The cooling rate calculated in this way was \unit[$(19 \pm 3)$]{$10^{3}\;\mathrm{s}^{-1}$}.

%-----------------------------
%-----------------------------

For both the nine-ion crystal and the eighteen-ion crystal, the fast cooling rates are a significant improvement over the ones obtained by sideband cooling on the quadrupole transition where several milliseconds of cooling time are required to cool multiple modes to close to the motional ground state \cite{Jurcevic2016}.

%-----------------------------
\begin{figure*}[hbt] \centering
\includegraphics[width=.75\linewidth]{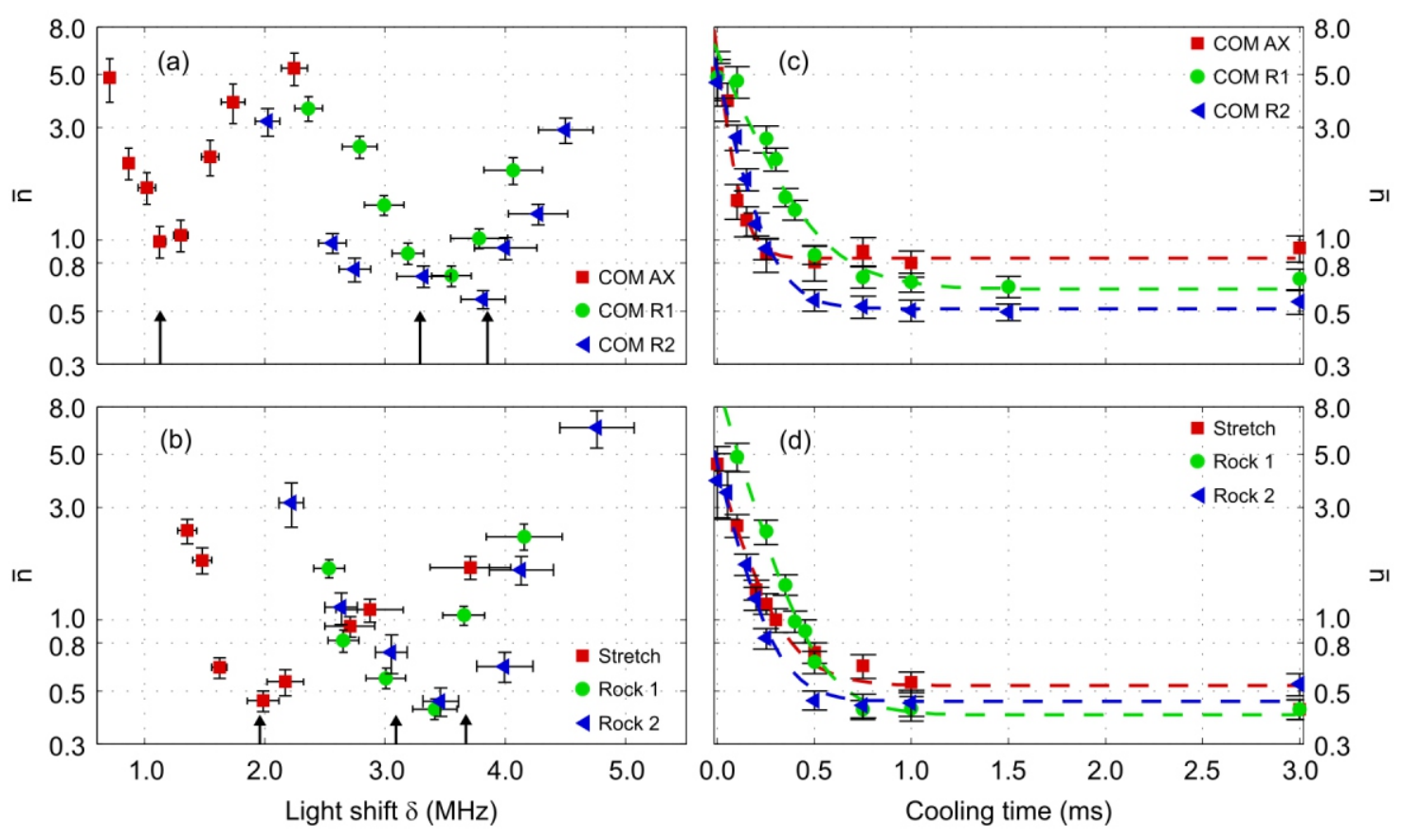}
%
% DATA: two-ion EIT cooling: 20140117
\caption{ \label{fig:dataseg} EIT cooling results of a two-ion crystal in setup 2. (a) Mean phonon numbers of the COM modes at \unit[$\left\{1.13, 3.29, 3.84\right\}$]{MHz}. (b) Mean phonon numbers of the stretch and rocking modes at \unit[$\left\{1.96, 3.09, 3.67\right\}$]{MHz}. Each data point in (a) and (b) consists of three measurements. One measurement is used to determine the light shift induced by the \mbox{$\sigma^{+}$-polarized} coupling laser. Two further measurements of Rabi oscillations on the red-sideband transition and blue-sideband transition each determine the mean phonon number by a thermal population distribution fit. (c) Cooling rates for COM modes. (d) Cooling rates for stretch and rocking modes. The mean phonon number was extracted using a simultaneous thermal distribution fit to the Rabi oscillations. All modes were cooled to the motional ground state within \unit[0.5]{ms}. Dashed lines represent the exponential fit to the data used to extract the cooling rate.} 
\end{figure*}	
%-----------------------------	

%%%%%%%%%%%%%%%%%%%%%%%%%%%%%%%%%%%%
\subsection{\label{ssub:segres} EIT cooling of a two-ion crystal in a segmented trap}

In the segmented trap setup, we investigated the EIT cooling range and rate for a two-ion crystal held in a potential
with center-of-mass mode frequencies of \unit[$\left\{\omega_{z},\omega_{r1},\omega_{r2}\right\}=2 \pi \left\{1.13, 3.29, 3.84\right\} $]{MHz}. We measured the mean phonon number of all six vibrational modes by analysis of Rabi oscillations on the red (blue) sidebands of the $\mathrm{S}_{1/2}\Leftrightarrow \mathrm{D}_{5/2}$ transition. These oscillations were recorded by coherently exciting only one of the ions using a strongly focused beam addressing one of the ions (with an intensity ratio $I_\mathrm{addr}/I_\mathrm{unaddr}<10^{-3}$). The mean phonon number was then determined by fitting the resulting sideband Rabi oscillations with a thermal phonon number distribution \cite{Meekhof1996}.

Figure\:\ref{fig:dataseg} depicts all data taken: the cooling efficiency is quantified in terms of the final phonon populations versus the induced EIT light shift (panels a,b), and in terms of achievable cooling rates (panels c,d). The upper panels display results for the COM modes;  the lower panels summarize the results for the stretch and rocking modes. While the cooling of the axial COM mode  reached a minimum phonon number of about one, all other modes were cooled to mean phonon numbers below one over a frequency range of about \unit[1]{MHz}.

The EIT cooling rate measurements shown in Fig.\:\ref{fig:dataseg}\:(c) and (d) were obtained for light shifts achieving the lowest mode temperatures. Exponential fits (indicated by dashed lines) were used to extract the cooling rate for the various modes, which are summarized in Tab.\:\ref{tab:segCR}. The measured cooling rates illustrate the potential for fast cooling with the EIT method, reaching a mean phonon number of below one within \unit[0.5]{ms} for all modes. This is a four-fold speed-up compared to the sideband cooling scheme commonly used in this experiment, where cooling a single mode to less than one phonon takes \unit[2]{ms} or more.

%-----------------------------
\begin{table}[htb]
\caption{\label{tab:segCR}Cooling rates in the segmented trap obtained from the fits shown in Fig.\:\ref{fig:dataseg}\:(c) and (d), and the respective mode frequencies.}
\begin{ruledtabular}
\begin{tabular}{cc}
Mode frequency (MHz)&Cooling rate ($10^{3}\, \mathrm{s}^{-1}$)\\
\hline
1.13& $17 \pm 7$\\
1.96 & $7 \pm 2$\\
3.08 & $7 \pm 1$\\
3.29 & $5 \pm 1$\\
3.67 & $9 \pm 2$\\
3.84 & $9 \pm 2$\\
\end{tabular}
\end{ruledtabular}
\end{table}
%-----------------------------

The observed minimum phonon number for COM modes were slightly higher than the ones of the out-of-phase modes. This effect is attributed to the larger COM-mode heating at rates that are only an order of magnitude smaller than the cooling rates in this setup. However, motional heating by stray electric fields is not the dominant mechanism limiting the lowest achievable mean phonon numbers. We attribute the worse cooling results observed in this setup (as compared to setup 1) to the less ideal cooling-beam geometry and to problems with achieving good micromotion compensation.

%%%%%%%%%%%%%%%%%%%%%%%%%%%%%%%%%%%%%%%%%%%%%%%%%%%%
\section{\label{sec:conc}Discussion and conclusion}

We investigated the use of EIT cooling for preparing the radial vibrational modes of long ion strings close to the ground state. Our experiments demonstrate that EIT cooling can indeed ground-state cool all modes over a wide frequency range, with sub-millisecond cooling times. This makes EIT cooling an attractive technique for trapped-ion experiments designed for studies of long-range Ising spin models. 

In recent experiments \cite{Jurcevic2014}, we cooled radial modes of long strings to the ground state by a series of sideband cooling pulses on a narrow transition, with each pulse being tailored to cool groups of modes in a particular frequency range. EIT cooling not only substantially shortens the cooling time as compared to sideband cooling on a narrow transition but also provides a more robust cooling technique with a smaller number of optimization parameters. 

We did observe, however, that in the currently used cooling configuration for highly anisotropic traps, EIT cooling gave rise to a substantial heating of low-frequency axial modes as the effective cooling $k$-vector also has an overlap with these modes. This is in contradiction to the predictions of the simple EIT cooling model presented in section \ref{sec:theo}. According to that model, all axial modes could be cooled below the Doppler limit achievable on the $\mathrm{S}_{1/2}\Leftrightarrow \mathrm{P}_{1/2}$ transition. This problem might be related to the experimental observation that the axial motion of the ion string was not in the Lamb-Dicke regime after Doppler cooling (as is assumed in the cooling model). We expect the ion motion along the axial direction to break up the coherent population trapping induced by the EIT beams, leading to additional light scattering and giving rise to heating effects. In future work, it would be interesting to measure the photon scattering rate during EIT cooling to infer the upper state populations for a comparison with the EIT cooling model. The difficulties described above could be overcome in future experiments by choosing an EIT cooling beam configuration featuring an EIT $k$-vector perpendicular to the ion string, so that it does not couple to and thereby heat low-frequency axial modes.

In conclusion, EIT cooling is an attractive technique for trapped-ion experiments requiring many vibrational modes to be cooled close to the ground states, as is the case in current quantum simulation experiments with long ion strings. Another interesting application would be the cooling of mixed-species ion crystals in quantum logic spectroscopy experiments where low motional energies are required for precision measurements of transition frequencies \cite{Rosenband2008}.

\begin{acknowledgments}
This work was supported by the Austrian Science Fund (FWF) under the grant number P25354-N20, by the European Commission via the integrated project SIQS and the ERC grant CRYTERION, and by the Institut f\"ur Quanteninformation GmbH.
\end{acknowledgments}

%\bibliography{bibliographyEIT}% Produces the bibliography via BibTeX.

\end{document}